\documentclass[multicol,showpacs,aps,nofootinbib]{revtex4}

\usepackage{graphicx}
\usepackage{feynmf}
\usepackage{psfig}

\newcommand{\la}{\left\langle}
\newcommand{\ra}{\right\rangle}
\newcommand{\be}{\begin{equation}}
\newcommand{\ee}{\end{equation}}
\newcommand{\bea}{\begin{eqnarray}}
\newcommand{\eea}{\end{eqnarray}}
\newcommand{\ba}{\begin{array}}
\newcommand{\ea}{\end{array}}
\newcommand{\piulug}{$\Pi^{u<}_{u>}$}
\newcommand{\piulbg}{$\Pi^{u<}_{b>}$}
\newcommand{\piulbl}{$\Pi^{u<}_{b<}$}
\newcommand{\piblbg}{$\Pi^{b<}_{b>}$}
\newcommand{\piblug}{$\Pi^{b<}_{u>}$}

\newcommand{\eqnMHDubk}
{\bea 
\left( -i\omega + \nu k^2 \right) u_i (\hat{k})  & 
 =  & -\frac{i}{2} P^+_{ijm}({\bf k}) \int d\hat{p} 
          [ u_j (\hat{p}) u_m (\hat{k}-\hat{p}) -
           b_j (\hat{p}) b_m (\hat{k}-\hat{p}) ] \label{eqn:udot} \\
\left( -i\omega + \eta k^2 \right) b_i (\hat{k})  & 
 = &  - i P^-_{ijm}({\bf k}) \int d\hat{p} 
          [ u_j (\hat{p}) b_m (\hat{k}-\hat{p}) ]
       		\label{eqn:bdot}   \\
k_i u_i({\bf k}) & = & 0 \\
k_i b_i({\bf k}) & = & 0  \eea}

\begin{document}
\title{Energy fluxes in helical magnetohydrodynamics and dynamo action}
\author{Mahendra\ K.\ Verma  \thanks{email: mkv@iitk.ac.in}}
\affiliation{Department of Physics, Indian Institute of Technology,
Kanpur  --  208016, INDIA}
\date{October 19, 2002}

\begin{abstract}
Renormalized viscosity, renormalized resistivity, and various energy
fluxes are calculated for helical magnetohydrodynamics using
perturbative field theory.  The calculation is to first-order in
perturbation.  Kinetic and magnetic helicities do not affect the
renormalized parameters, but they induce an inverse cascade of
magnetic energy.  The sources for the the large-scale magnetic field
have been shown to be (1) energy flux from large-scale velocity field
to large-scale magnetic field arising due to nonhelical interactions,
and (2) inverse energy flux of magnetic energy caused by helical
interactions. Based on our flux results, a premitive model for
galactic dynamo has been constructed. Our calculations yields dynamo
time-scale for a typical galaxy to be of the order of $10^8$ years.
Our field-theoretic calculations also reveal that the flux of magnetic
helicity is backward, consistent with the earlier observations based
on absolute equilibrium theory.

\end{abstract}
\vspace{1.5cm}
\maketitle 

\section{Introduction}
Generation of magnetic field in plasma, usually referred to as
``dynamo'', is one of the prominent and unsolved problems in physics
and astrophysics.  It is known that the magnetic field of galaxies,
the Sun, and the Earth are neither due to some permanent magnet nor
due to any remnants of the past, but it is generated by the nonlinear
processes of plasma motion (\cite{Moff:book,Krau:book} and references
therein).  However, a solid quantitative understanding is lacking in
this area inspite of various attempts for more than half century.
There are various aspects in this problem, and we address energy
transfer issues in this paper and in paper I (\cite{MKV:MHDflux})
using field-theoretic methods in a somewhat idealized environment,
homogeneous and isotropic flows.  In paper I we show that the
nonhelical part of magnetohydrodynamic (MHD) interaction causes energy
transfer from large-scale (LS) velocity field to large-scale (LS)
magnetic field.  In a typical dynamo environment however, the helical
interactions cause an additional energy cascade of magnetic energy
from small scales (SS) to large scales; the field-theoretic
calculation of helical contribution to the energy flux is presented in
this paper.  Both helical and nonhelical factors contribute to the
magnetic energy growth.

In the problem of magnetic field generation, it is required that the
LS magnetic field is maintained at all time.  There are
several exact results in this area, e.g., dynamo does not exist in two
dimensions as well as in axisymmetric flows \cite{Moff:book}.  In the
past, several dynamos, e.g., rotor dynamo, 2-sphere dynamo etc., have
been constructed \cite{Moff:book}, however, mean-field electrodynamics
 developed by Steenbeck et al.  \cite{Stee} (also see Krause and
R\"{a}dler \cite{Krau:book}) paved a way for practical calculations in
astrophysical and terrestrial dynamo.  This formalism also provided
insights into the physical mechanism of dynamo, mainly that kinetic
helicity $H_K = 1/2 (\la {\bf u \cdot \Omega} \ra$, where ${\bf u}$ and
${\bf \Omega}$ are the velocity and vorticity fields respectively)
plays an important role in the amplification of the magnetic field.
The amplification parameter $\alpha_u$ was found to be $(\tau/3) H_K$,
where $\tau$ is the velocity de-correlation time.

Mean-field electrodynamics of Steenbeck et al.~\cite{Stee} is a
kinematic theory. Here it is assumed that the velocity field is a
known function, which is unaffected by the generated magnetic field.
The later models which take into account the back reaction of the
magnetic field to the velocity field are called dynamic models.  One
of the first dynamic model is due to Pouquet et al.~\cite{Pouq:EDQNM}
where they incorporated the feedback, and proposed that the modified
$\alpha$ is proportional to residual helicity, $H_K-H_J$, where $H_J$
is the current helicity define as $1/2 \la {\bf b \cdot \nabla b} \ra$.  
Gruzinov and Diamond \cite{Diam:dyna}
proposed a quenching mechanism where the amplification parameter was
modified to
\begin{equation}
\alpha = \frac{\alpha_u}{1+ R_m (\overline{B}/\overline{U})^2} 
\end{equation}
where $\overline{B}^2$ is the LS magnetic energy,
$\overline{U}$ is the LS velocity field, and $R_m$ is the
magnetic Reynolds number.  Recently Field, Blackman, and Chou
\cite{Fiel}, and Chou \cite{Chou:theo} obtained a general expression
for dynamic $\alpha$ coefficient as a function of Reynolds number and
magnetic Prandtl number.  Basu and Bhatthacharya \cite{Basu1} and
Basu \cite{Basu2} have attempted to compute the dynamo coefficients
$\alpha$ and $\beta$ using field-theoretic techniques.

In a recent development Brandenburg \cite{Bran:num} investigated
dynamo problem in isotropic and helical MHD.  The the system is
forced with kinetic helicity, they find magnetic energy transfer to
large-scales.  They indentify this mechanism, named as nonlinear
alpha-effect, for magnetic energy growth.  Magnetic helicity play
an important role in this mechanism; we will discuss these issues
in later part of the paper.

There are many numerical simulations of dynamo in various geometries.
In simulations MHD equations are numerically solved with appropriate
boundary conditions.  In this paper we will only refer to the
simulations performed for a periodic box; this is to avoid the
complications of spherical geometry (e.g., effects of Coriolis force
etc.).  Pouquet et al.~\cite{Pouq:EDQNM} numerically integrated the
MHD equations on the basis of Eddy-damped-quasi-normal-Markovian
(EDQNM) approximation.  When kinetic energy (KE=$1/2 \la {\bf u \cdot u}
\ra$) and $H_K$ were injected near a wavenumber band, both magnetic
energy (ME=$1/2 \la {\bf b \cdot b} \ra$) and absolute value of
magnetic helicity $|H_M|$ were found to increase.  Magnetic helicity
$H_M$ is defined as $ 1/2 \la {\bf a \cdot b}\ra $, where ${\bf a}$ is
vector potential.  Note however that one part of the time-scale used
in Pouquet et al.'s calculation is based on Alfv\'{e}n relaxation
time.  This assumption is suspect in view of current theoretical
\cite{Srid1,Srid2,MKV:B0RG,MKV:MHD_PRE,MKV:MHDRG} and numerical
results \cite{MKV:MHDsimu,Bisk1:Kolm,Bisk2:Kolm}, which favour
$k^{-5/3}$ (Kolmogorov's) energy spectrum over Kraichnan's $k^{-3/2}$
spectrum for MHD turbulence.  In our present paper, the nonlinear
time-scale is based on Kolmogorov's energy spectrum.  The direct
numerical simulation (DNS) of Pouquet and Patterson \cite{Pouq:num}
yielded a similar result.  The results of Brandenburg
\cite{Bran:num} discussed above are also obtained numerically.
Frick and Sokoloff \cite{Fric} studied the shell model of turbulence,
and showed that magnetic helicity suppresses turbulent cascade.

In another development Dar et al.~\cite{Dar:flux} numerically
calculated various energy fluxes in two-dimensional MHD and showed
that there is significant energy transfer from LS velocity field to LS
magnetic field; they claimed that the above flux is one of the main
contributor to the amplification of large-scale ME.  Note that
the ratio ME/KE grows in both helical and nonhelical MHD as well as 
in many decaying as well as forced simulations (see for
example, \cite{Dar:flux,Pouq:nonhelical} and references therein).
Hence, helicity is not a necessary requirement for the generation of
magnetic energy.  Regarding flux of magnetic helicity, Pouquet et
al.~\cite{Pouq:EDQNM} and Pouquet and Patterson \cite{Pouq:num} argue
that it is in the inverse direction (from SS to LS).  We find a
similar magnetic helicity flux in our theoretical calculation.

In one of the recent theoretical development Kulsrud and Anderson
\cite{Kuls1} derived and solved the kinetic equation for the growth of
galactic magnetic field.  They argued that the dynamo time-scale is
much large than the growth time-scale of turbulent modes.  Hence, the
buildup of SS turbulent ME dominates the slow growth of LS ME, thus
making the mean dynamo theory invalid.

In paper I and the present paper, the energy fluxes of MHD turbulence
are computed using perturbative field theory.  The calculation is to
first order in perturbation.  Here the viscosity and resistivity were
taken from renormalization calculation of Verma
\cite{MKV:MHD_PRE,MKV:MHDRG}, and the energy spectrum were taken to
Kolmogorov-like ($k^{-5/3}$).  Kolmogorov's spectrum for MHD
turbulence is supported by recent theoretical
\cite{Srid1,Srid2,MKV:B0RG,MKV:MHD_PRE,MKV:MHDRG} and numerical
results \cite{MKV:MHDsimu,Bisk1:Kolm,Bisk2:Kolm}.  In paper I we show
that in a kinetically forced nonhelical MHD, the energy transfer from
LS velocity field to LS magnetic field is one of the dominant
transfers.  It is also shown in paper I that the above energy flux
into LS magnetic modes is independent of the nature of LS forcing.
In this paper we generalize the field-theoretic
calculation of Verma \cite{MKV:MHD_PRE,MKV:MHDflux} to helical MHD.
As discussed in paper I, we assume that the turbulence is homogeneous
and isotropic, and that the mean magnetic field is absent.  The
absence of mean magnetic field is a reasonable assumption for the
initial stages of dynamo evolution.  This assumption is to ensure that
the turbulence is isotropic.  In addition we take cross helicity $( 
{\bf u \cdot b})$ to be zero to simplify the calculation.  We examine
the fluxes in the presence of magnetic and kinetic helicities.  We
also investigate the flux of magnetic helicity.

We have constructed a simple dynamo model using the theoretically-calculated
energy flux into the large-scale magnetic field.  In this model the ME
grows exponentially in the initial stage, and the growth time-scale is
of the order of eddy turnover time.  It shows that dynamo action is
possible in galactic dynamo.

The outline of the paper is as follows: in section 2, we carry out the
perturbative calculation of renormalized viscosity and resistivity, as
well as fluxes of energy and magnetic helicity.  In section 3, we
construct a dynamic galactic dynamo based on our energy flux results.
In this section we also compare our results with the findings of
earlier researchers.  Section 4 contains conclusions.

\section{Field-theoretic calculation of helical magnetohydrodynamics}
The incompressible MHD equation in Fourier space is given by
\eqnMHDubk
where ${\bf u}$  and  ${\bf b}$  are  the velocity and magnetic  field
fluctuations respectively, $\nu$ and $\eta$  are the viscosity and the
resistivity respectively, and
\begin{eqnarray}
P^{+}_{ijm}({\bf k}) & = & k_j P_{im}({\bf k}) + k_m P_{ij}({\bf k});
\\ P_{im}({\bf k}) & = & \delta_{im}-\frac{k_i k_m}{k^2}; \\
P^-_{ijm}({\bf k}) & = & k_j \delta_{im} - k_m \delta_{ij}; \\
\hat{k}  =  ({\bf k},\omega); & \hspace{1cm} &
d \hat{p}  = d {\bf p} d \omega/(2 \pi)^{4}.
\end{eqnarray}
Note that we are working in three space dimension and also with
zero mean magnetic field.

Some of the definitions regarding kinetic and magnetic helicities are
in order.  Throughout this paper, ${\bf a}$ denotes the vector
potential (${\bf b = \nabla \times a}$), and ${\bf \Omega}$ denotes
the vorticity (${\bf \Omega = \nabla \times u}$).  The spectrum of
helicity, $H_M({\bf k})$, is defined using the equal-time correlation
function $\la a_i({\bf k},t) b_j({\bf k},t) \ra$ (the angular brackets
denote ensemble average),
\begin{equation}
\la a_i({\bf k},t) b_j({\bf k'},t) \ra  
\stackrel{\mathrm{def}}{=}
P_{ij}({\bf k}) H_M({\bf k})   (2 \pi)^3 \delta({\bf k+k'})
\end{equation}
The factor $P_{ij}({\bf k})$ appears due to the constraints
$\nabla \cdot {\bf a} = \nabla \cdot {\bf b}=0$.  Using this
correlation function we derive the following relationship:
\begin{eqnarray}
H_M & = & \frac{1}{2} \la {\bf a(x) \cdot b(x)} \ra  \nonumber \\
    & = & \frac{1}{2} \int \frac{d{\bf k}}{(2 \pi)^3}
                           \frac{d{\bf k'}}{(2 \pi)^3}
                           \la {\bf a(k) \cdot b(k')} \ra  \nonumber \\ 
    & = & \int \frac{d{\bf k}}{(2 \pi)^3} H_M({\bf k})
\end{eqnarray}
The one-dimensional magnetic helicity $H_M(k)$ is defined using
\begin{equation}
\int H_M(k) dk = \int \frac{d{\bf k}}{(2 \pi)^3} H_M({\bf k})
\end{equation}
Therefore,
\begin{equation}
H_M(k) = \frac{4 \pi k^{2}}{(2 \pi)^3} H_M({\bf k})
\end{equation}
Using $\nabla \times {\bf a = b}$ we can easily derive that
\begin{equation}
{\bf a(k)} = \frac{i}{k^2} {\bf k \times b(k)}
\end{equation}
which leads to
\begin{equation}
\label{eqn:bibj}
\la b_i ({\bf k}) b_j ({\bf k'}) \ra  =  
         \left[ P_{ij}({\bf k)} C^{bb} ({\bf k}) -
           i \epsilon_{ijl} k_l  H_M(k) \right] (2 \pi)^3 \delta({\bf k+k'}) 
\end{equation}
where $C^{bb}$ is the $b-b$ correlation function.

A similar analysis for kinetic helicity shows that
\begin{equation}
\la v_i({\bf k}) \Omega_j({\bf k'}) \ra  
\stackrel{\mathrm{def}}{=} 
 P_{ij}({\bf k}) H_K({\bf k})  \delta({\bf k+k'})
\end{equation}
\begin{equation}
H_K  = \frac{1}{2} \la {\bf u \cdot \Omega} \ra =
	\int \frac{d{\bf k}}{(2 \pi)^3} H_K({\bf k})
\end{equation}
and
\begin{equation}
\label{eqn:uiuj}
\la u_i ({\bf k}) u_j ({\bf k'}) \ra   = 
        \left[ P_{ij}({\bf k}) C^{uu} ({\bf k}) -
       i \epsilon_{ijl} k_l \frac{ H_K(k)}{k^2}  \right] 
	(2 \pi)^3 \delta({\bf k+k'}) 
\end{equation}
where $C^{uu}$ is the $u-u$ correlation function.

An important point to note is that magnetic helicity is conserved in
MHD when $\nu=\eta=0$.  One of the consequences of this
conservation law is the emergence of $k^{-1}$ spectrum at smaller
wavenumbers \cite{FrisPouq}.  The kinetic helicity is conserved in
fluid turbulence, but not in MHD turbulence.

In the following subsection we will calculate the renormalized
viscosity and resistivity for helical MHD.

\subsection{Calculation of  renormalized parameters}

Recently Verma \cite{MKV:MHD_PRE,MKV:MHDRG} has calculated
the renormalized viscosity and resistivity for MHD turbulence
in absence of kinetic and magnetic helicity.   It will
be shown below that the presence of both 
kinetic and magnetic helicities does not alter the renormalized
viscosity and resistivity calculated for nonhelical MHD.

In the RG procedure the wavenumber range $(k_N,k_0)$ is divided
logarithmically into $N$ shells. Then the elimination of the first
shell $k^>=(k_1,k_0)$ is carried out and the modified MHD equation for
$k^<=(k_{N},k_1)$ is obtained.  This process is continued for higher 
shells. The shell elimination is performed by ensemble averaging over
$k^>$ modes \cite{McCo:book,MKV:MHD_PRE,MKV:MHDRG}.  It is assumed
that $u_i^>(\hat{k})$ and $b_i^{>}(\hat{k})$ have gaussian
distributions with zero mean, while $u_i^{<}(\hat{k})$ and
$b_i^{<}(\hat{k})$ are unaffected by the averaging process.  In
addition it is also assumed that
\begin{eqnarray}
\la u_i^> (\hat{p}) u_j^> (\hat{q}) \ra  & = &
\left[ P_{ij}({\bf p)} C^{uu} (\hat{p}) -
i \epsilon_{ijl} \frac{p_l}{p^2} H_K(\hat{p}) \right] (2 \pi)^4
\delta(\hat{p}+\hat{q}) 
\label{eqn:avgbegin}\\
\la b_i^> (\hat{p}) b_j^> (\hat{q}) \ra  &= &
\left[ P_{ij}({\bf p)} C^{bb} (\hat{p}) -
i \epsilon_{ijl} p_l  H_M (\hat{p}) \right]  (2 \pi)^4
\delta(\hat{p}+\hat{q}) \label{eqn:avgend}
\end{eqnarray}
Note that $u-b$ correlation has been taken to be zero in our calculation.

We apply first-order perturbation theory to compute the renormalized
parameters.  After elimination of $n$ shells, we obtain the  
following equations for the renormalized viscosity $\nu_{(n)}$
and renormalized resistivity $\eta_{(n)}$ (for details refer
to Verma \cite{MKV:MHDRG}).
\begin{eqnarray}
\left( -i\omega + \nu_{(n)} k^2 + 
	\delta \nu_{(n)} k^2 \right) u_i^<(\hat{k}) & = & 
	-\frac{i}{2} P^+_{ijm}({\bf k}) \int d\hat{p}
	[u_j^< (\hat{p}) u_m^< (\hat{k}-\hat{p}) \nonumber \\ 
& & \hspace{1in} - b_j^<(\hat{p}) b_m^< (\hat{k}-\hat{p}) ] \\
\left( -i\omega + \eta_{(n)} k^2 
	+ \delta \eta_{(n)} k^2 \right) b_i^<(\hat{k}) & = &  
	-i P^-_{ijm}({\bf k}) \int d\hat{p} 
	[u_j^< (\hat{p}) b_m^< (\hat{k}-\hat{p}) ] 
\end{eqnarray} 
where
\begin{eqnarray}
\delta \nu_{(n)}(k) & = & \frac{1}{2 k^2} 
			  \int^{\Delta}_{\hat{p}+\hat{q}=\hat{k}}
			    \frac{d {\bf p}}{(2 \pi)^3}
                            \left[ \frac{S(k,p,q) C^{uu}(q)+S'(k,p,q) H_K(q)} 
 			         {\nu_{(n)}(p) p^2+\nu_{(n)}(q) q^2} \right.
			       \nonumber \\   
 		    & & \hspace{1.5in} 
                       \left. - \frac{S_6(k,p,q)C^{bb}(q)+S'_6(k,p,q) H_M(q)}
			      {\eta_{(n)}(p) p^2+\eta_{(n)}(q) q^2} \right]
                         \label{eqn:nu} \\  \nonumber \\
\delta \eta_{(n)}(k) & = & \frac{1}{2 k^2}
			      \int^{\Delta}_{\hat{p}+\hat{q}=\hat{k}}
				 \frac{d {\bf p}}{(2 \pi)^3}
			   \left[-\frac{S_8(k,p,q)C^{bb}(q)+S'_8(k,p,q)H_M(q)}
			      {\nu_{(n)}(p) p^2+\eta_{(n)}(q) q^2} \right.
				\nonumber \\   
                        &  & \hspace{1.5in} 
                         \left. +\frac{S_9(k,p,q)C^{uu}(q)+S'_9(k,p,q)H_k(q)}
			     {\eta_{(n)}(p) p^2+\nu_{(n)}(q) q^2} \right]
                             \label{eqn:eta}
\end{eqnarray}
The quantities $S_i$ and $S'_i$ are as follows:
\begin{eqnarray}
S(k,p,q)   & = & P^{+}_{bjm}(k) P^{+}_{mab}(p) P_{ja}(q) 
             =  2 kp \left( z^3+xy \right) \\ 
S_6(k,p,q) & = & P^{+}_{ajm}(k) P^{-}_{mba}(p) P_{jb}(q)
             =  - 2 kp z \left( 1- y^2 \right) \\
S_8(k,p,q) & = & P^{-}_{ijm}(k) P^{+}_{jab}(p) P_{ma}(q) P_{ib}(k) 
             =  S_6(p,k,q) \\
S_9(k,p,q) & = & P^{-}_{ijm}(k) P^{-}_{mab}(p) P_{ja}(q) P_{ib}(k) 
             =  2 kp(z+xy) \\
S'(k,p,q)  & = & P^{+}_{bjm}(k) P^{+}_{mab}(p) \epsilon_{jal} q_l =0 \\
S'_6(k,p,q) & = & P^{+}_{ajm}(k) P^{-}_{mba}(p) \epsilon_{jal} q_l =0 \\
S'_8(k,p,q) & = & P^{-}_{ijm}(k) P^{+}_{jab}(p) \epsilon_{mal} q_l
						P_{ib}(k) = 0   \\
S'_9(k,p,q) & = & P^{-}_{ijm}(k) P^{-}_{mab}(p) \epsilon_{jal} q_l
						P_{ib}(k) = 0
\end{eqnarray}
Since $\delta \nu$ and $\delta \eta$ are proper scalars and $H_{M,K}$
are pseudo scalars, $S'_i(k,p,q)$ will be pseudo scalars.  In
addition, $S'_i(k,p,q)$ are also linear in $k,p$ and $q$.  This
implies that $S'_i(k,p,q)$ must be proportional to ${\bf q \cdot (k
\times p)}$, which will be zero because ${\bf k=p+q}$.  Hence all
$S'_i(k,p,q)$ turn out to be zero, as a consequence the presence of
helicities does not alter the already calculated $\delta
(\nu,\eta)_{(n)} (k)$ by Verma \cite{MKV:MHD_PRE,MKV:MHDRG}.
Zhou \cite{Zhou:HK} arrived at a similar conclusion while calculating
the renormalized viscosity for helical fluid turbulence.

Verma \cite{MKV:MHD_PRE,MKV:MHDRG} obtained a self-consistent solution
of the renormalized parameters using Kolmogorov's spectrum.  Since the
helicities do not alter the renormalized parameters, we arrive at the
same formula for renormalized viscosity and resisitivity as Verma
\cite{MKV:MHD_PRE,MKV:MHDRG}, that is,
\begin{eqnarray}
\label{eqn:nu_eta_k}
(\nu,\eta)(k) & = & \left\{ \begin{array}{ll}
	(K^u)^{1/2} \Pi^{1/3} k^{-4/3} (\nu^*,\eta^*)    
	& \mbox{\hspace{1cm} for $k \ge k_n$}  \\
 	(K^u)^{1/2} \Pi^{1/3} k_n^{-4/3} (\nu^*,\eta^*)    
	& \mbox{\hspace{1cm} for $k \le k_n$} 
			      \end{array}
	               \right.
\end{eqnarray}
where $\Pi$ is the total energy flux, $K^u$ is the Kolmogorov's
constant, and $\nu^*$ and $\eta^*$ are the renormalized parameters.
The value of these renormalized parameters have been listed in
\cite{MKV:MHD_PRE,MKV:MHDRG}.

The present calculation has been carried out up to first order.  The
probability distribution of velocity is gaussian, while that of
velocity difference is nongaussian.  The nongaussian behaviour of
velocity difference has significant effects specially on higher order
structure functions, which are not properly accounted for by first
order calculations. Yet, the first order calculation of renormalized
viscosity yields results very close to those obtained in experiments
and numerical simulations (see e.g., \cite{YakhOrsz,McCo:book}).  For
the above reason, we have stuck to the first-order field-theoretic
calculation in the present paper.

In the next subsection we will calculate the energy and helicity
fluxes  using the field theoretic technique. 

\subsection{Calculation of energy and helicity fluxes}

In paper I we have analytically calculated energy fluxes in the
absence of magnetic and kinetic helicities. In this subsection we will
generalize that calculation for helical MHD.  Refer to paper I for the
energy evolution equations and other basic formulas.

As discussed in paper I, the energy flux from inside of the $X$-sphere
($X<$) to outside of the $Y$-sphere ($Y>$) is 
\begin{eqnarray}
\Pi^{X<}_{Y>}(k_0) & = & \frac{1}{(2 \pi)^3 \delta({\bf k'+p+q})}
                        \int_{k'>k_0} \frac{d {\bf k'}}{(2 \pi)^3} 
		       \int_{p<k_0} \frac{d {\bf p}}{(2 \pi)^3}  
			\la S^{YX}({\bf k'|p|q}) \ra
\label{eqn:flux}		
\end{eqnarray} 
where $X$ and $Y$ stand for $u$ or $b$, and $S({\bf k'|p|q})$ is
energy transfer from mode {\bf p} of $X$ field to mode {\bf k} of $Y$
field, with mode {\bf q} acting as a mediator.  The detailed expressions
for $\la S^{YX}({\bf k'|p|q}) \ra$ are given in Paper I.

We calculate the above fluxes analytically to the leading order in
perturbation series using the same procedure as in paper I.  Some
additional terms appear in $ \la S(k'|p|q) \ra$ due to the presence of
helicity.  The detailed expressions are given in Appendix A.

A formula for the magnetic helicity flux can be derived in a similar
manner.  From Eqs.~(\ref{eqn:udot},\ref{eqn:bdot}) we can easily
obtain the equation for the evolution of magnetic helicity, which is
\begin{eqnarray}
\frac{\partial H_M({\bf k})}{\partial t} & = & \frac{1}{2 (2 \pi)^3 \delta({\bf k+k'})}
         \Re \left[{\bf b^*(k)} \cdot \frac{\partial {\bf a(k')}}{\partial t}
            +{\bf a^*(k)} \cdot \frac{\partial {\bf b(k')}}{\partial t}\right] \\
            & = & \frac{1}{2 (2 \pi)^3 \delta({\bf k+k'})}
                  \left[ S^{H_M}({\bf k'|p|q})+S^{H_M}({\bf k'|q|p})\right]
\end{eqnarray}
where
\begin{eqnarray}
S^{H_M}({\bf k'|p|q}) & = & \frac{1}{4} \Re\left[{\bf b(k')
                        \cdot (v(p) \times b(q))}\right] \nonumber \\
        &   &+\frac{1}{4} \Im\left[
              {\bf k' \cdot b(q) a(k') \cdot u(p)
                 - k' \cdot u(q) a(k') \cdot b(p)} \right]
\end{eqnarray}
Here $\Re()$ and $\Im()$ stand for real and imaginary part of the
arguments, respectively.  The quantity $S^{H_M}({\bf k'|q|p})$ can be
obtained from the above expression by interchanging ${\bf p}$ and
${\bf q}$.  After some algebraic manipulation it can be shown that
\begin{eqnarray}
S^{H_M}({\bf k'|p|q}) + S^{H_M}({\bf k'|q|p}) + 
S^{H_M}({\bf p|k'|q}) & &   \nonumber \\
+ S^{H_M}({\bf p|q|k'}) + 
S^{H_M}({\bf q|k'|p}) + 
S^{H_M}({\bf q|p|k'}) & = & 0
\end{eqnarray}
It shows that the ``detailed conservation of magnetic helicity''
holds in a triad interaction (when $\nu=\eta=0$) \cite{Lesl:book}.

From the above, the transfer rate of magnetic helicity from a wavenumber sphere
of radius $k_0$ is
\begin{equation}
\Pi_{H_M}(k_0) = \frac{1}{(2 \pi)^3 \delta({\bf k'+p+q})}
                    \int_{k'>k_0} \frac{d {\bf k'}}{(2 \pi)^3} 
		       \int_{p<k_0} \frac{d {\bf p}}{(2 \pi)^3}  
                        \la S^{H_M}({\bf k'|p|q}) \ra
\label{eqn:HMflux}         
\end{equation}
We again compute $\la S^{H_M}({\bf k'|p|q}) \ra$ to first order in
perturbation.  The detailed expressions are given in Appendix B.

The expressions in the Appendices involve Green's functions and
correlation functions.  The expressions for these functions are taken
from self-consistent calculations (see e.g., Verma \cite{MKV:MHDRG}).
For $G(k,t-t')$ of the formulas (\ref{eqn:Suu}-\ref{eqn:Sbb},\ref{eqn:SHm}), we
substitute 
\begin{equation}
G^{(uu,bb)}(k,t-t')  =  \theta(t-t') \exp{\left[-\left( \nu(k),\eta(k) \right) 
				k^2 (t-t')\right]}
\label{eqn:G}
\end{equation}
where $(\nu(k),\eta(k))$ are given by Eq.~(\ref{eqn:nu_eta_k}), and 
$\theta(t-t')$ is the step function.  We
assume the relaxation time-scale for $C^{uu}(k,t,t')$ and
$H_K(k,t,t')$ to be $(\nu(k) k^2)^{-1}$, while that of
$C^{bb}(k,t,t')$ and $H_M(k,t,t')$ to be $(\eta(k) k^2)^{-1}$.  The
spectrum $C^{(uu,bb)}(k,t,t)$ are written in terms of one-dimensional
energy spectra $E^{(u,b)}$ as
\begin{equation}
C^{(uu,bb)}(k,t,t) = \frac{(2 \pi)^3}{4 \pi k^2} E^{(u,b)}
\end{equation}

In presence of magnetic helicity, the calculations based on absolute
equilibrium theories suggest that the energy cascades forward, and the
magnetic helicity cascades backward \cite{FrisPouq}. In this paper
we have not considered the inverse cascade region of magnetic
helicity. We take Kolmogorov's spectrum for energy
based on recent numerical simulations
\cite{MKV:MHDsimu,Bisk1:Kolm,Bisk2:Kolm} and theoretical calculations
\cite{Srid1,Srid2,MKV:B0RG,MKV:MHDRG} (ignoring the intermittency
corrections).    Hence, the spectrum of $E^{(u,b)}$ can be taken as
\begin{eqnarray}
\label{eqn:Ek_helical}
E^u(k) & = & 	K^u \Pi^{2/3} k^{-5/3} \\
E^b(k) & = & E^u /r_A
\label{eqn:energy_spectra}
\end{eqnarray}
where $\Pi$ is total energy flux.

The helicities are written in terms of energy spectra as
\begin{eqnarray}
\label{eqn:HK}
H_K(k) & = & r_K k E^u(k) \\
H_M(k) & = & r_M \frac{E^b(k)}{k}
\label{eqn:HM}
\end{eqnarray} 
We are calculating energy fluxes for the inertial-range wavenumbers
where the same powerlaw is valid for all energy spectrum.
Therefore, the ratios $r_A, r_M$, and $r_K$ can be treated as
constants. 

We substitute the above forms for the correlation and Green's
functions [Eqs.~(\ref{eqn:G}-\ref{eqn:energy_spectra})] in the
expressions for $\la S^{YX}(k'|p|q)\ra$ and $\la S^{H_M}(k'|p|q) \ra$
given in the Appendices.  These $S$'s are substituted in the flux
formulas (Eqs.~[\ref{eqn:flux},
\ref{eqn:HMflux}]).  We  make the following change of variable:
\begin{equation}
k=\frac{k_0}{u}; p=\frac{k_0}{u} v; q=\frac{k_0}{u} w
\end{equation}
These operations yield the following nondimensional form of the
equation in the $-5/3$ region (for details, refer to
\cite{MKV:MHDflux}).
\begin{eqnarray}
\frac{\Pi^{X<}_{Y>}(k_0)}{\Pi(k_0)} & = & 
	(K^u)^{3/2} \left[ \frac{1}{2}
         \int_0^1 dv \ln{(1/v)} \int_{1-v}^{1+v} dw 
		 (vw) \sin \alpha F^{X<}_{Y>} \right]
\label{eqn:piE_ratio1} \\  \nonumber \\
\frac{\Pi_{H_M}(k_0)}{\Pi(k_0)} & = & \frac{1}{k_0}
	(K^u)^{3/2} \left[ \frac{1}{2}
         \int_0^1 dv (1-v) \int_{1-v}^{1+v} dw 
		 (vw) \sin \alpha F_{H_M} \right]
\label{eqn:piHm_ratio1}
\end{eqnarray}
where the integrands $(F^{X<}_{Y>},F_{H_M})$ are function of $v$, $w$,
$\nu^*$, $\eta^*$, $r_A, r_K$ and $r_M$ \cite{MKV:MHDflux}. 

We compute the term in the square brackets, $I^{X<}_{Y>}$, using the
similar procedure as that of Verma \cite{MKV:MHDflux}.  
The flux ratios $\Pi^{X<}_{Y>}/\Pi$ can be written in terms of
integrals $I^{X<}_{Y>}$, which have been computed numerically.  Table
\ref{tab:rA} contains their values for $r_A=1$
and $r_A=5000$. The constant
$K^u$ is calculated using the fact that the total energy
flux $\Pi$ is sum of all $\Pi^{X<}_{Y>}$.  For parameters ($r_A=5000,
r_K=0.1, r_M=-0.1$), $K^u=1.53$, while for ($r_A=1, r_K=0.1,
r_M=-0.1$), $K^u=0.78$.  After this the energy flux ratios
$\Pi^{X<}_{Y>}/\Pi$ can be calculated.  These ratios for some of the
specific values of $r_A$, $r_K$ and $r_M$ are listed in Table
\ref{tab:flux_ratio}.  The first and second terms of $\Pi^{X}_{Y}/\Pi$
entries are nonhelical and helical components respectively.

An observation of the results shows some interesting patterns.  The
energy flux can be split into two parts: helical (dependent on $r_K$
and/or $r_M$) and nonhelical (independent of helicity).  The
nonhelical part of all the fluxes except \piblug
($\Pi^{b<}_{u>nonhelical} < 0$ for $r_A > 1$) is always positive.  As
a consequence, in nonhelical channel, ME cascades from LS to SS.
Also, since $\Pi^{u<}_{b<} >0$, LS kinetic energy feeds the LS
magnetic energy.  The fluxes of nonhelical MHD has been discussed in
great detail in paper I.

The sign of $\Pi^{u<}_{u>helical}$ is always negative, i.e., kinetic
helicity reduces the KE flux.  But the sign of helical
component of other energy fluxes depends quite crucially on the sign
of helicities. From the entries of Table \ref{tab:rA}, we see that
\begin{equation}
\Pi^{b<}_{(b>,u>)helical} = - a r_M^2+b r_M r_K
\label{eqn:rM_dependence}
\end{equation}
where $a$ and $b$ are positive constants.  If $r_M r_K <0$, the energy
flux to LS magnetic field due to both the terms in the right-hand-side
of the above equation is positive.  Earlier EDQNM \cite{Pouq:EDQNM}
and numerical simulations \cite{Bran:num} with forcing of KE and $H_K$
typically have $r_K r_M <0$.  Hence, we can claim that helicity
typically induces an inverse energy cascade via $\Pi^{b<}_{b>}$ and
$\Pi^{b<}_{u>}$.  These fluxes will enhance the large-scale magnetic
field.

From the entries of Table~\ref{tab:flux_ratio} we can
infer that the for small and moderate $r_K$ and $r_M$, the
inverse energy cascade into large-scale
magnetic field is less than the forward 
nonhelical energy flux $\Pi^{\b<}_{b>}$.  While for helical
MHD ($r_K,r_M \rightarrow 1$), the inverse helical cascade
dominates the nonhelical magnetic-to-magnetic energy cascade.

The flux ratio $\Pi_{H_M}/\Pi$ can be written in terms the integrals
of Eqs.~(\ref{eqn:piHm_ratio1},\ref{eqn:piHm_ratio1}) using the same
procedure as done for energy flux ratios.  The numerical values
of the integrals are shown in Tables 1 and 2.  Clearly, 
\begin{equation}
\Pi_{H_M} = -d r_M + e r_K
\label{eqn:Hm_flux}
\end{equation}
where $d$ and $e$ are positive constants.  Note however that
contribution of $H_M$ dominates that of $H_K$.  Clearly the sign of
$\Pi_{H_M}$ is the same as that of $H_K$ but negative of $H_M$.  From
the above equation we observe that positive $H_M$ yields a negative
contribution to $\Pi_{H_M}$. Hence, for positive $H_M$, the magnetic
helicity cascade is backward.  This result is in agreement with Frisch
et al.'s \cite{FrisPouq} argument in which they predict an inverse
cascade of magnetic helicity.  Our theoretical result on inverse
cascade of $H_M$ is also in agreement with the results derived using
EDQNM calculation \cite{Pouq:EDQNM} and numerical simulations
\cite{Pouq:num}.  

When we force the system with positive kinetic helicity ($r_K>0$),
Eq.~(\ref{eqn:Hm_flux}) indicates a forward cascade of magnetic
energy. This effect could be the reason for the observe production of positive
magnetic helicity at small scale by Brandenburg \cite{Bran:num}
 Because of magnetic helicity conservation, 
he also finds generation of negative magnetic helicity at large-scales.
Now, positive kinetic helicity and negative magnetic helicity 
at large-scales may yield an inverse cascade of magnetic
energy (see Eq.~[\ref{eqn:rM_dependence}).  This could be the
crude reason for the growth of magnetic energy in the simulations
of Brandenburg \cite{Bran:num}.

In paper I we calculated $\Pi^{u<}_{b<}$ for nonhelical MHD using
steady-state condition. 
\begin{equation}
\Pi^{u<}_{b<} = \Pi^{b<}_{b> nonhelical} + \Pi^{b<}_{u> nonhelical} .
\label{eqn:piulbl}
\end{equation}
 The above calculation for helical MHD is not
straight forward because magnetic energy at large-scale could increase
with time, and steady state may not be achievable for all possible
parameters of helical MHD.  Brandenburg~\cite{Bran:num} observes
the dynamic evolution of large-scale magnetic energy in his
simulations.  To simplify the calculation, we assume steady-state
condition for helical MHD as well, and calculate various parameters.
For some set of highly helical MHD, we get negative energy flux.

In the following section, we will construct a dynamic dynamo
model for galaxies using our flux results.

\section{Dynamo via energy and magnetic helicity fluxes}

In the above calculation we have assumed that the turbulence is
homogeneous, isotropic, and steady.  The assumption of homogeneity and
isotropy can be assumed to hold in galaxies in the early phases of
evolution before large structures appear.  The assumption that the
mean magnetic field of galaxy is rather small is valid in the
beginning of the galactic evolution. Therefore, we apply the flux
obtained from our calculations to estimate the growth of
magnetic energy in galaxies.

During the early phase of galactic evolution, only the large-scales
(LS)  contain the
kinetic and magnetic energies.  The fields at these scales interact
with each other, but the small-scale spectrum is far from steady (not
enough time).  The interactions of LS velocity field and the LS seed
magnetic field increase the LS seed magnetic field $E^b(t)$ till the
steady-state is reached.  In absence of helicity, the source of energy
for the large-scale magnetic field is \piulbl
[Eq.~(\ref{eqn:piulbl})].  When helicity is present, there are several
other sources as discussed in section 2 of this paper.  Since the
forcing of helicities is effective at LS, it is reasonable to assume
that the helical part of $\Pi^{b>}_{b<}$ and $\Pi^{u>}_{b<}$ will also
aid to the increase in LS ME.  Hence,
\begin{equation}
\label{eqn:dynamo_Ebdot}
\frac{d E^b(t)}{dt} =  \Pi^{u<}_{b<}+ \Pi^{b>}_{b<helical} + 
			\Pi^{u>}_{b<helical}
\end{equation}
We assume a quasi-steady approximation for the early evolution of
magnetic field. In many quasi-steady situations (slowly decaying or
growing), steady-state results are usually applied.  This
approximation works very well for many practical problems.  We make
this assumption in this paper, and substitute the theoretically
calculated energy fluxes calculated in Section II of the paper to the
above equation.

Since the ME starts with a small value
(large $r_A$ limit), all the fluxes appearing in
Eq.~(\ref{eqn:dynamo_Ebdot}) are proportional to $r_A^{-1}$ [cf.
Eqs.~(\ref{eqn:Sub},\ref{eqn:Sbb})], i.e.,
\begin{equation}
\Pi^{u<}_{b<} + \Pi^{b>}_{b<helical} + \Pi^{u>}_{b<helical}
		= c \Pi \frac{E^b}{E^u}
\end{equation}
where $E^u$ is the LS KE, and $c$ is the constant of proportionality,
which depends on the values of helicities.  Both Kinetic and magnetic
helicities are difficult to ascertain for a galaxy due to lack of
observations.  We take $r_M$ and $r_K$ to be of the order of 0.1, with
$r_M$ being negative.  The choice of negative $r_M$ is motivated by
the results of EDQNM calculation \cite{Pouq:EDQNM} and numerical
simulations \cite{Bran:num}.  With this value of $r_M$ and $r_K$, $c
\approx 0.84$ for $E(k) \propto k^{-5/3}$ regime, and $c \approx 1.3$
for $E(k) \propto k^{-1}$ regime. Since both the values of constant
$c$ is approximately equal and close to 1.0, we take $c=1.0$ for our
calculation.  Hence,
\begin{eqnarray}
 \frac{1}{\Pi} \frac{d E^b}{d t} & 
	\approx &   \frac{E^b}{E^u}
\end{eqnarray}
Using $E^u = K^u \Pi^{2/3} L^{2/3}$, where $L$ is the large length-scale of
the system, we obtain
\begin{equation}
 \frac{1}{\sqrt{E^u} E^b} \frac{d E^b}{d t} 
	\approx \frac{1}{L (K^u)^{3/2}}
\end{equation}
We assume that $E^u$ does not change appreciably in the early phase.
Therefore,
\begin{equation}
E^b(t) \approx E^b(0) \exp{\left(\frac{ \sqrt{E^u}}{L (K^u)^{3/2}} 
		t \right)}
\end{equation}
Hence, the ME grows exponentially in the early periods, and the
time-scale of growth is of the order of $L (K^u)^{3/2}/\sqrt{E^u}$,
which is the eddy turnover time \cite{MKV:MHDflux}.  Taking $L \approx
10^{17} km$ and $\sqrt{E^u} \approx 10 km/sec$, we obtain the growth
time-scale to be $10^{16} sec$ or $3 \times 10^8$ years, which is in
the expected range \cite{Kuls1}.  Hence, we have constructed a
nonlinear and dynamically consistent galactic dynamo based on the
energy fluxes.  In this model the ME grows exponentially, and the
growth time-scale is reasonable \cite{Kuls1}.  

The helical and nonhelical contribution to the fluxes for $r_A=5000,
r_K=0.1, r_M=-0.1$ is shown in Table~\ref{tab:flux_ratio}.  The flux
ratios shown in the table do not change appreciably as long as $r_A >
100$ or so.  The three fluxes responsible for
the growth of LS ME are $\Pi^{u<}_{b<}/\Pi \approx 2.6 \times 10^{-4}$
(nonhelical), $\Pi^{b<}_{b>helical}/\Pi \approx -4.1 \times 10^{-5}$,
and $\Pi^{b<}_{u>helical}/\Pi \approx -4.0 \times 10^{-5}$.  The ratio
of nonhelical to helical contribution is $2.6/0.81 \approx 3.2$.
Hence, the nonhelical contribution is significant, if not more, than
the contribution from the helical part for the LS ME amplification.
Note that in the earlier papers on dynamo, the helical part is
strongly emphasized.

Kulsrad and Anderson (KA) \cite{Kuls1} performed an important mean
field dynamo calculation of galactic dynamo for large Prandtl numbers.
Some of the salient features are follows.  In KA's kinematic dynamo
calculation the growth rate of ME is $\gamma = (1/3) \int k^2 U({\bf
k}) d{\bf k}$ where
\begin{equation}
U({\bf k}) = \frac{(\Pi)^{2/3} k^{-5/3}}{4 \pi k^3 u_k}
	  \approx (\Pi)^{1/3} k^{-13/3}
\end{equation}
KA estimate $\gamma \approx k_{max}^{2/3} (\Pi)^{1/3}
\approx 10^{-4} yr^{-1}$.  From this result KA conclude that the
kinematic theory predicts a extremely rapid growth of SS ME.
The SS noisy magnetic field thus generated will dominate the
mean magnetic field that grows at a considerably slower rate (dynamo
growth time $\approx 3 \times 10^8 yr$).  Therefore, it is claimed
that the kinematic assumption of the mean dynamo theory is invalid,
and it is difficult to build up galactic magnetic field from a very
weak seed field using dynamo action.  
KA's estimate of growth time-scale is
equal to the eddy turnover time of smallest eddies ($k_{max}^{-1}$).
  Hence, as pointed out by KA,
kinematic assumption is invalid for galactic dynamo, and one has to
resort to a dynamical model.  Brandenburg's numerical results 
\cite{Bran:num} are not quite consistent with KA's results.
For example, Brandenburg finds (1) growth of magnetic energy
even for large magnetic Prandtl number; (2) the growth time-scale
for magnetic energy is of the order of $L^2/\eta$, where
$L$ is the large length-scale, and $\eta$ is resistivity.

Our results are valid for Prandtl number close to 1.  Therefore,
they can not be compared with KA's calculations. It is interesting
to see however that our crude estimate of time scale is one eddy turn-over
time.  To get a better picture, we need to construct a more
solid model.

In our model the magnetic energy growth is due to the fluxes \piulbl +
$\Pi^{b>}_{b<helical}$ + $\Pi^{u>}_{b<helical}$.  In nonhelical MHD,
only \piulbl is effective, while in helical MHD both kinetic and
magnetic helicities play an important role in the growth of ME.  In
the current kinematic models of planetary magnetism \cite{Moff:book},
magnetic field is generated by kinetic helicity, for which planetary
rotation (spin) plays an important role.  These models appear to work
for all the planets except Mercury, which rotates far too slowly.  We
conjecture that \piulbl   (independent of helicities) probably plays an
important role in the generation of magnetic field of Mercury.

In helical MHD, the helical contribution to the magnetic energy growth
goes as [see Eq.~(\ref{eqn:rM_dependence})]
\begin{equation}
\frac{d E^b}{d t} = a r_M^2 - b r_M r_K
\label{eqn:Ebdot2}
\end{equation}
where $a$ and $b$ are positive constants.  The term
$a r_M^2$ is always positive independent of the sign of $H_M$, but
$-b r_M r_K$ is positive only when  $H_M$ and $H_K$ are of the opposite
sign.  In numerical simulation of Brandenburg \cite{Bran:num} and
EDQNM calculation of Pouquet et al.~\cite{Pouq:EDQNM}, $H_M H_K <0$
for small $k$, and $H_M H_K > 0$ for large $k$.  Hence, $-b r_M r_K$
term is positive for small $k$, resulting in positive $dE^b /dt$.
Let us compare the above result with the dynamical dynamo of
Pouquet et al.~\cite{Pouq:EDQNM}, Field et al.~\cite{Fiel},
and Cho \cite{Chou:theo}.

The kinematic dynamo predicts that the growth parameter $\alpha$ is
proportional to $H_K$, i.e, $\alpha = \alpha_u \propto \la {\bf u \cdot
\nabla \times u} \ra$.  The kinetic model was generalized by Pouquet et
al.~\cite{Pouq:EDQNM}, Field et al.~\cite{Fiel}, and Cho \cite{Chou:theo}.
In absence of a mean magnetic field they find that 
\begin{equation}
\alpha \approx \alpha_u + \alpha_b = 
	... \la {\bf u \cdot \nabla \times u} \ra
	- ... \la {\bf b \cdot \nabla \times b} \ra,
\label{eqn:Pouq} 
\end{equation}
where $...$ denotes certain time scales (which are always
positive).  It implies that $\alpha$ gets positive contribution from
both the terms when $H_M$ and $H_K$ are of opposite signs.  This
result is consistent with Eq.~(\ref{eqn:Ebdot2}).  The direct
numerical simulation of Pouquet and Patterson \cite{Pouq:num} indicate
that $H_M$ enhances the growth rate of ME considerably, but that is
not the case with $H_K$ alone.  This numerical result is somewhat
inconsistent with results of Pouquet et al.~and others \cite{Pouq:EDQNM}
(Eq.~(\ref{eqn:Pouq})), but it fits better our formula
(\ref{eqn:Ebdot2}) ($d E^b/dt=0$ if $r_M=0$).  Hence, our formula
(\ref{eqn:Ebdot2}) probably is a better model for the dynamically
consistent dynamo.

In the following section we will summarize our results.

\section{Conclusions}

In this paper we have applied first-order perturbative field theory to
calculate the renormalized viscosity, renormalized resistivity, and
various cascade rates for helical MHD.  We find that the renormalized
viscosity and resistivity are unaffected on introduction of both
kinetic and magnetic helicities.  Our result is consistent with Zhou's
calculation \cite{Zhou:HK} for helical fluid turbulence.

We find that the energy cascade rates get significantly altered by
helicity.  Since magnetic helicity is a conserved quantity in MHD,
Frisch \cite{FrisPouq} had argued for $k^{-1}$ energy spectra at small
wavenumbers. However, in this paper we calculate energy fluxes in the
Kolmogorov's inertial range, where we find direct energy cascades.
The fluxes are shown in Table~\ref{tab:flux_ratio}.  The main results
of our calculation are as follows:

\begin{enumerate}
\item The magnetic energy flux has two components: (a) the nonhelical
part which is always positive, (b) the helical part which is negative
(assuming $H_M H_K <0$).  The inverse cascade resulting due to
helicities is consistent with the results of Pouquet et
al.~\cite{Pouq:EDQNM}, Brandenburg \cite{Bran:num}, and others.

\item The $u-u$ flux \piulug   gets a inverse component due to kinetic
helicity.  This implies that KE flux decreases in presence of $H_K$, a
result consistent with that of Kraichnan \cite{Krai:HK}.  

\item The growth of large-scale magnetic field in the initial stage of
evolution results from
\piulbl+$\Pi^{b>}_{b<helical}+\Pi^{u>}_{b<helical}$.  In this paper we
have computed the relative magnitudes of all three contributions, and
find that all of them to be comparable, although $\Pi^{u<}_{b<}$ is
somewhat higher.  Pouquet et al.~\cite{Pouq:EDQNM}, Pouquet and
Patterson \cite{Pouq:num}, Brandenburg \cite{Bran:num}, and many
others highlight $\Pi^{b>}_{b<helical}$ transfer, and generally do not
consider \piulbl and $\Pi^{u>}_{b<helical}$ fluxes.

\item Regarding positive $H_M$, the flux of magnetic helicity
$\Pi_{H_M}$ is backward.

\end{enumerate}

Most of the earlier papers (e.g. Pouquet et al.~\cite{Pouq:EDQNM})
assume Alfv\'{e}n time-scale to be the dominant time-scale for MHD
turbulence.  We have taken nonlinear time-scale (based on Kolmogorov's
spectrum) to be the relevant time-scale based on recent numerical
\cite{MKV:MHDsimu,Bisk1:Kolm,Bisk2:Kolm} and theoretical
\cite{Srid1,Srid2,MKV:B0RG,MKV:MHD_PRE,MKV:MHDRG} work.

Using the flux results we have constructed a simple nonlinear and dynamically
consistent galactic dynamo.  Our model shows an exponential growth of
magnetic energy in the early phase (much before saturation).  The
growth time-scale is of the order of $3 \times 10^8$ years, which is
consistent with the current estimate \cite{Kuls3}.
In our paper we have discussed the growth of magnetic field at
scale comparable to forcing scales.  In real dynamo,
the magnetic field at even larger scales also grow.
This growth may be due to inverse cascade of magnetic
energy.  This problem is beyond the scope of our paper.

Some of the results presented here are general, and they are expected
to hold in solar and planetary dynamo.  For example, we find that
LS velocity field supply energy to LS seed magnetic
field.  This is one of the sources of dynamo.  Hence, if we solve first
few MHD modes in spherical coordinate with kinetic forcing, it may be
possible to capture some of the salient features of solar and
planetary dynamo.

In summary, the energy flux studies of helical MHD provide us with
many important insights into the problem of magnetic energy growth.
Its application to galactic dynamo yields very interesting results.  A
generalization of the formalism presented here to spherical geometry
may provide us with insights into the magnetic field generation in the
Sun and Earth.

\acknowledgements 
The author thanks G. Dar, V. Eswaran, A. Brandenburg, D. Narsimhan,
and R. K. Varma for discussions.  He also thanks Mustansir Barma
(TIFR, Mumbai) and Krishna Kumar (ISI, Calcutta) for useful
suggestions and kind hospitality during his stay in their institutes
on his sabbatical leave. This work was supported in part by Department
of Science and Technology, India.

\appendix
\section{Calculation of $\la S^{YX}({\bf k'|p|q}) \ra$}

The expressions
for $\la S^{YX} \ra$ for helical MHD are
\begin{eqnarray}
\label{eqn:Suu}
\la S^{uu}(k'|p|q)\ra & = & \int_{-\infty}^t  dt'  (2 \pi)^3 \left[ 
                 T_1(k,p,q) G^{uu}(k,t-t') C^{u}(p,t,t') C^{u}(q,t,t') 
				\right. \nonumber \\
& &\hspace{1cm} +T'_1(k,p,q) G^{uu}(k,t-t') \frac{H_K(p,t,t')}{p^2} 
				\frac{H_K(q,t,t')}{q^2} \nonumber \\
& &\hspace{1cm} +T_5(k,p,q)  G^{uu}(p,t-t') C^{u}(k,t,t') C^{u}(q,t,t') 
							\nonumber \\
& &\hspace{1cm}	+T'_5(k,p,q) G^{uu}(p,t-t') \frac{H_K(k,t,t')}{k^2} 
				\frac{H_K(q,t,t')}{q^2} \nonumber \\	
& &\hspace{1cm} +T_9(k,p,q)  G^{uu}(q,t-t') C^{u}(k,t,t') C^{u}(p,t,t') 
							\nonumber \\
& &\hspace{1cm}	\left. +T'_9(k,p,q)G^{uu}(q,t-t') \frac{H_K(k,t,t')}{k^2} 
				\frac{H_K(p,t,t')}{p^2} \right] \\
-\la S^{ub}(k'|p|q) \ra & = & \int_{-\infty}^t  dt' (2 \pi)^3 \left[
  		 T_{2}(k,p,q) G^{uu}(k,t-t') C^{b}(p,t,t') C^{b}(q,t,t') 
						\right. \nonumber \\
& &\hspace{1cm} +T'_{2}(k,p,q)G^{uu}(k,t-t') H_M(p,t,t') H_M(q,t,t')	
							\nonumber \\
& &\hspace{1cm} +T_{7}(k,p,q) G^{bb}(p,t-t') C^{u}(k,t,t') C^{b}(q,t,t') 
							\nonumber \\
& &\hspace{1cm} +T'_{7}(k,p,q)G^{bb}(p,t-t') \frac{H_K(k,t,t')}{k^2} 
					H_M(q,t,t')     \nonumber \\	
& &\hspace{1cm} +T_{11}(k,p,q)G^{uu}(q,t-t') C^{u}(k,t,t') C^{b}(p,t,t') 
							\nonumber \\
& &\hspace{1cm} \left. +T'_{11}(k,p,q)G^{uu}(q,t-t') \frac{H_K(k,t,t')}{k^2} 
					H_M(p,t,t') \right] 
					\label{eqn:Sub} \\
-\la S^{bu}(k'|p|q) \ra & = & \int_{-\infty}^t  dt' (2 \pi)^3 \left[
                 T_{3}(k,p,q) G^{bb}(k,t-t') C^{u}(p,t,t') C^{b}(q,t,t') 
					\right. \nonumber \\
& &\hspace{1cm}	+T'_{3}(k,p,q)G^{bb}(k,t-t') \frac{H_K(p,t,t')}{p^2} 
				     H_M(q,t,t') \nonumber \\
& &\hspace{1cm} +T_{6}(k,p,q) G^{uu}(p,t-t') C^{b}(k,t,t') C^{b}(q,t,t') 
						\nonumber \\
& &\hspace{1cm}	+T'_{6}(k,p,q) G^{uu}(p,t-t') H_M(k,t,t') H_M(q,t,t') 
						\nonumber \\
& &\hspace{1cm} +T_{12}(k,p,q)G^{bb}(q,t-t') C^{b}(k,t,t') C^{u}(p,t,t') 
						\nonumber \\
& &\hspace{1cm} \left. +T'_{12}(k,p,q)G^{bb}(q,t-t') H_M(k,t,t') 
				\frac{H_K(p,t,t')}{p^2} \right] 
					\label{eqn:Sbu} \\
\la S^{bb}(k'|p|q) \ra & = & \int_{-\infty}^t  dt' (2 \pi)^3 \left[
                 T_{4}(k,p,q) G^{bb}(k,t-t') C^{b}(p,t,t') C^{u}(q,t,t') 
						\right.	\nonumber \\
& &\hspace{1cm} +T'_{4}(k,p,q)G^{bb}(k,t-t') H_M(p,t,t') 
				\frac{H_K(q,t,t')}{q^2} \nonumber \\
& &\hspace{1cm} +T_{8}(k,p,q) G^{bb}(p,t-t') C^{b}(k,t,t') C^{u}(q,t,t') 
							\nonumber \\
& &\hspace{1cm} +T'_{8}(k,p,q)G^{bb}(p,t-t') H_M(k,t,t') 
				\frac{H_K(q,t,t')}{q^2} \nonumber \\
& &\hspace{1cm} +T_{10}(k,p,q)G^{uu}(q,t-t') C^{b}(k,t,t') C^{b}(p,t,t') 
							\nonumber \\
& &\hspace{1cm}	\left. +T'_{10}(k,p,q)G^{uu}(q,t-t') H_M(k,t,t') H_M(p,t,t') 
						\right] 
\label{eqn:Sbb}
\end{eqnarray}
where $T_i (k,p,q)$ are given in paper I.  To obtain $T'_i(k,p,q)$
(helical part) we replace all the second rank tensors of the type
$P_{ja}(k)$ by $\epsilon_{jal}k_l$.

\section{Calculation of $\la S^{H_M}({\bf k'|p|q}) \ra$}

The quantity $\la S^{H_M}({\bf k'|p|q}) \ra$ 
of Eq.~(\ref{eqn:HMflux}) simplifies to
\begin{eqnarray}
\la S^{H_M}({\bf k'|p|q}) \ra & = & \frac{1}{2} \Re \left[ 
\epsilon_{ijm} \la b_i(k') u_j(p) b_m(q) \ra  \right. \nonumber \\
& & - \epsilon_{jlm} \frac{k_i k_l}{k^2} 
		\la u_i(q) b_m(k') b_j(p) \ra  \nonumber \\
& & \left. +\epsilon_{jlm} \frac{k_i k_l}{k^2} 
		\la b_i(q) b_m(k') u_j(p) \ra \right],
\end{eqnarray}
which is computed perturbatively to the first order.  The corresponding
Feynman diagrams are
\input{hmfeyn.diag}

Here empty, shaded, and filled triangles (vertices) represent
$\epsilon_{ijm}, - \epsilon_{ijm} k_i k_l /k^2$ and $\epsilon_{ijm}
k_i k_l/k^2$ respectively.  The empty and filled circles (vertices)
denote $(-i/2)P^-_{ijm}$ and $-iP^+_{ijm}$ respectively.  The solid,
dashed, wiggly (photon), and curly (gluons) lines denote $\la u_i u_j
\ra, \la b_i b_j \ra, G^{uu}$, and $G^{bb}$ respectively.  When we
substitute $\la u_i u_j \ra, \la b_i b_j \ra$ using
Eqs.~(\ref{eqn:uiuj},\ref{eqn:bibj}), we obtain terms involving
$C^{X}(p,t,t') H_{(M,K)}(q,t,t')$. The resulting expression for $\la
S^{H_M}({\bf k'|p|q}) \ra$ is
\begin{eqnarray}
\label{eqn:SHm}
\la S^{H_M}({\bf k'|p|q})\ra & = & \int_{-\infty}^t  dt' (2 \pi)^3 \left[ 
	 T_{31}(k,p,q) G^{bb}(k,t-t') \frac{H_K(p,t-t')}{p^2} 
		C^b(q,t-t') \right. \nonumber \\
& &\hspace{1cm} +T_{32}(k,p,q) G^{bb}(k,t-t') C^{uu}(p,t-t') H_M(q,t-t') \nonumber \\
& &\hspace{1cm} +T_{33}(k,p,q) G^{uu}(p,t-t') H_M(k,t-t') C^{bb}(q,t-t') \nonumber \\
& &\hspace{1cm} +T_{34}(k,p,q) G^{uu}(p,t-t') C^{bb}(k,t-t') H_M(q,t-t') \nonumber \\  
& &\hspace{1cm} +T_{35}(k,p,q) G^{bb}(q,t-t') H_M(k,t-t') C^{uu}(p,t-t') \nonumber \\
& &\hspace{1cm} +T_{36}(k,p,q) G^{bb}(q,t-t') C^{bb}(k,t-t') 
			\frac{H_K(p,t-t')}{p^2}   \nonumber \\
& &\hspace{1cm} +T_{37}(k,p,q) G^{bb}(k,t-t') H_M(p,t-t') C^{uu}(q,t-t') \nonumber \\
& &\hspace{1cm} +T_{38}(k,p,q) G^{bb}(k,t-t') C^{bb}(p,t-t') \frac{H_K(q,t-t')}{q^2} 
								      \nonumber \\
& &\hspace{1cm}	+T_{39}(k,p,q) G^{bb}(p,t-t') H_M(k,t-t') C^{uu}(q,t-t') \nonumber \\
& &\hspace{1cm} +T_{40}(k,p,q) G^{bb}(p,t-t') C^{bb}(k,t-t') \frac{H_K(q,t-t')}{q^2} 
								      \nonumber \\ 
& &\hspace{1cm} +T_{41}(k,p,q) G^{uu}(q,t-t') H_M(k,t-t') C^{bb}(p,t-t') \nonumber \\
& &\hspace{1cm} +T_{42}(k,p,q) G^{uu}(q,t-t') C^{bb}(k,t-t') H_M(p,t-t') \}\nonumber \\
& &\hspace{1cm} +T_{43}(k,p,q) G^{bb}(k,t-t') \frac{H_K(p,t-t')}{p^2} C^{bb}(q,t-t') 
								      \nonumber \\
& &\hspace{1cm} +T_{44}(k,p,q) G^{bb}(k,t-t') C^{uu}(p,t-t') H_M(q,t-t') \}\nonumber \\
& &\hspace{1cm} +T_{45}(k,p,q) G^{uu}(p,t-t') H_M(k,t-t') C^{bb}(q,t-t') \nonumber \\
& &\hspace{1cm} +T_{46}(k,p,q) G^{uu}(p,t-t') C^{bb}(k,t-t') H_M(q,t-t') \nonumber \\
& &\hspace{1cm} +T_{47}(k,p,q) G^{bb}(q,t-t') H_M(k,t-t') C^{uu}(p,t-t') \nonumber \\
& &\hspace{1cm} \left. +T_{48}(k,p,q) G^{bb}(q,t-t') C^{bb}(k,t-t') 
		\frac{H_K(p,t-t')}{p^2} \right]
\end{eqnarray}
The terms $T_{31..48}(k,p,q)$ can be obtained in
terms of antisymmetric tensors $\epsilon_{jal}$, $P_{ijm}^{\pm}$ etc.
They have not been listed here due to lack of space.


\pagebreak

\begin{table}
\caption{The values of $I^X_Y= (\Pi^{X}_{Y}/\Pi)/(K^u)^{1.5}$ calculated
using Eqs.~(\ref{eqn:piE_ratio1}-\ref{eqn:piHm_ratio1}) 
for Alfv\'{e}n ratios $r_A=1$ and $r_A=5000$.}
\label{tab:rA}
\begin{ruledtabular} 
\begin{tabular}{lcc} 
              & $r_A=1$                     & $r_A=5000$   \\ \hline 
$I^{u<}_{u>}$ & $0.19-0.10 r_K^2$              &  $0.53-0.28 r_K^2$   \\
$I^{u<}_{b>}$ & $0.62+0.3 r_M^2+0.095 r_K r_M$ 
	      & $1.9 \times 10^{-4}+1.4 \times 10^{-9} r_M^2
			+2.1 \times 10^{-5} r_K r_M$ \\
$I^{b<}_{u>}$ & $0.18-2.04 r_M^2+1.93 r_K r_M$ 
	      & $-5.6\times 10^{-5}-1.1 \times 10^{-7} r_M^2
			+5.4\times 10^{-4} r_K r_M$ \\
$I^{b<}_{b>}$ & $0.54-1.9 r_M^2+2.02 r_K r_M$   
	      & $1.4 \times 10^{-4}-1.02 \times 10^{-7} r_M^2
			+5.4 \times  10^{-4} r_K r_M$  \\
$I^{u<}_{b<}$ & -    & - \\
$I_{H_M}$     & $-25 r_M + 0.35 r_K$      
	      & $-4.1 \times 10^{-3} r_M + 8.1 \times 10^{-5} r_K$
\end{tabular}
\end{ruledtabular}
\end{table}

\begin{table}
\caption{The values of energy flux ratios $\Pi^{X}_{Y}/\Pi$ for
various values of $r_A, r_K,$ and $r_M$ for $k^{-5/3}$ region.  The
first and second entries are nonhelical and helical contributions
respectively. }
\label{tab:flux_ratio}
\begin{ruledtabular} 
\begin{tabular}{lccccc} 
$(r_A,r_K,r_M)$  & \piulug/$\Pi$ & \piulbg/$\Pi$ &
\piblug/$\Pi$ & \piblbg/$\Pi$ \\ \hline

(5000,0.1,-0.1) & ($1.0,-0.0053$) & ($3.2 \times 10^{-4}$,  
   & ($-9.7 \times 10^{-7}$,
   & ($2.5 \times 10^{-4}$, \\
                &             & $-3.7 \times 10^{-7}$) 
   & $-9.0 \times 10^{-6}$)   & $-9.4 \times 10^{-6}$)  & \\ 

(5000,0.1,0.1) & ($1.0,-0.0053$)  & ($3.2 \times 10^{-4}$, 
   & ($-9.7 \times 10^{-5}$,
   & ($2.5 \times 10^{-4}$, \\
   &              &  $3.7 \times 10^{-7}$)
   & $9.0 \times 10^{-6}$)    & $9.4 \times 10^{-6}$)  & \\

(1,0.1,-0.1) & (0.13,$-6.9 \times 10^{-4}$) & (0.43,$-4.4 \times 10^{-4}$) 
             & ($0.13,-0.027$)
             & ($0.37,-0.027$)    \\

(1,0.1,0.1)  & (0.12,$-6.5 \times 10^{-4}$) & (0.4,$8.1 \times 10^{-4}$) 
	     & (0.12,$-7.7 \times 10^{-4}$
             & (0.35,$8.3 \times 10^{-4}$     \\

(1,1,-1) & ($0.029,-0.015$) & (0.095,$-9.9 \times 10^{-3}$) 
             & ($0.028,-0.61$)
             & ($0.083,-0.60$)  \\

(1,1,1) & ($0.12,-0.064$) & (0.39,0.079) & ($0.12,-0.075$) 
             & (0.34,0.081)  \\

(1,0,1) & (0.081,0) & (0.26,0.013) & ($0.078,-0.86$) 
            & ($0.23,-0.8$) 
 
\end{tabular}
\end{ruledtabular}
\end{table}

\end{document}